%% file: rare.tex
\documentclass{elsart}
\usepackage{graphicx}
\usepackage{amssymb}

\newcommand{\cd}{$^{116}$Cd}
\newcommand{\cdhdz}{$^{113}$Cd}
\newcommand{\cdhs}{$^{106}$Cd}
\newcommand{\cdha}{$^{108}$Cd}
\newcommand{\zns}{$^{70}$Zn}
\newcommand{\znvs}{$^{64}$Zn}
\newcommand{\tehz}{$^{120}$Te}

\newcommand{\tehaz}{$^{128}$Te}
\newcommand{\tehd}{$^{130}$Te}

\newcommand{\bdec}{$\beta$-decay}

\newcommand{\EC}{EC}
\newcommand{\qval}{$Q$-value}
\newcommand{\qvals}{$Q$-values}
\newcommand{\majo}{Majorana}
\newcommand{\onbb}{neutrinoless double beta decay}
\newcommand{\nel}{\mbox{$\nu_e$}}
\newcommand{\neu}{neutrino}
\newcommand{\neus}{neutrinos}
\newcommand{\obb}{0\mbox{$\nu\beta\beta$-decay}} 
\newcommand{\zbb}{2\mbox{$\nu\beta\beta$-decay}} 
\newcommand{\bmbm}{\mbox{$\beta^-\beta^-$}}
\newcommand{\bpbp}{\mbox{$\beta^+\beta^+$}}
\newcommand{\bec}{\mbox{$\beta^+/\mbox{EC}$}}
\newcommand{\ecec}{\mbox{EC/EC}} 
\newcommand{\ema}{\mbox{$\langle m_{\nu_e} \rangle$ }}

\newcommand{\nbb}{$0\nu\beta^-\beta^-$}

\newcommand{\npe}{$0\nu\beta^+\!\mbox{EC}$}

\newcommand{\npp}{$0\nu\beta^+\beta^+$}

\newcommand{\tee}{$2\nu\mbox{ECEC}$}

\newcommand{\keV}{ke$\!$V}

\newcommand{\Times}{\times}
\newcommand{\sDelta}{{\scriptstyle\Delta}}
\def\ra{\rightarrow}

\newcommand{\eg}{e.g.}
\newcommand{\fig}{figure}
\newcommand{\Fig}{Figure}
\newcommand{\tab}{table}
\newcommand{\tabs}{tables}

\def\thepage{\roman{page}}

\begin{document}
\begin{frontmatter}

\title{A Search for various Double Beta Decay Modes of Cd, Te and Zn Isotopes}
\author{H.~Kiel, D.~M\"unstermann}
\address{Lehrstuhl f\"ur Experimentelle Physik IV, Universit\"at Dortmund,\\
Otto--Hahn Str. 4, 44221 Dortmund, Germany}
\author{K.~Zuber}
\address{Denys Wilkinson Laboratory, Dept. of
Physics, University of Oxford,\\ Keble Road, Oxford, OX1 3RH, UK}

\begin{abstract}
Various double beta decay modes of Cd, Zn and Te isotopes are
explored with the help of CdTe and CdZnTe semiconductor detectors.
The data set is splitted in an energy range below 1\,MeV having a
statistics of 134.5\,g$\cdot$d and one above 1\,MeV resulting in
532\,g$\cdot$d. No signals were observed in all channels under investigation.
New improved limits for the neutrinoless double beta decay
of \zns\ of $T_{1/2} > 1.3 \cdot 10^{16} yrs$ (90\%\,CL), the longest standing
limit of all double beta isotopes, and
\npe\ of \tehz\ of $T_{1/2} > 2.2 \cdot 10^{16} yrs$ (90\%\,CL) are given. 
For the first time a limit on the half-life of the \tee\ of $^{120}$Te
of $T_{1/2} > 9.4 \cdot 10^{15} yrs$ (90\%\,CL) is obtained. In addition,  
limits on \tee\ for ground state transitions of \cdhs, \cdha\ and \znvs\ are improved.
The obtained results even under rough background conditions show the reliability
of CdTe semiconductor detectors for rare nuclear decay searches.
\end{abstract}
\end{frontmatter}

\section{Introduction}
\input{kai.tex}

\input{setup.tex}

\input{data.tex}

\input{analysis.tex}

\input{modes.tex}

\section{Summary}
The main intention of the paper is to show the reliability of
CdTe detectors for the search of rare nuclear decays, especially double
beta decay.
Within the context of background studies for the planned COBRA-experiment
\cite{COBRA}, a setup consisting of two CdTe/CdZnTe detectors
in a shielding was used to investigate various double beta decay
modes of Cd, Te and Zn isotopes.  The searches took advantage of
the fact that source and detector are identical and of the good 
energy resolution of semiconductor detectors.
With a collected statistics of 0.53\,kg$\cdot$d for energies beyond 1 MeV 
and 0.13\,kg$\cdot$d for energies below 1 MeV, limits of the order
of 10$^{19}\,yrs$ for various \nbb-modes could be obtained. A new
improved limit for the \obb\ of ${}^{70}$Zn is given, which has not
been investigated experimentally in the last fifty years.
A detailed
study of all available \bpbp -channels is performed, where for the
first time values for the \tee-decay of $^{120}$Te are obtained. Limits on the
\npe\ of $^{120}$Te and the \tee\ for ground state transitions of
\cdhs, \cdha\ and \znvs\ were improved significantly.\\
Having shown with the obtained results, that CdTe can be used for low-level applications,
because the material
seems to be clean enough that an upgrade to larger detector masses like 10 kg as 
proposed in \cite{COBRA} would allow to probe neutrino masses below 1 eV.
Especially in case of \cd{} it seems to be possible to make the experiment
background free. The main advantage is, that the half-life sensitivity scales with
$M^{1/2}$ instead of $M^{1/4}$ as in the background limited case \cite{steve}. 
The first advantage is the high Q-value of 2805 keV of \cd{} \obb{} which is 
above all gamma-lines occuring in the natural decay chains. Thus this 
normally dominating background component can be neglected. 
The only beta decay in the natural decay chains with an
endpoint higher than 2805 keV stems from $^{214}$Bi, but this potential background
can be easily vetoed because it is associated with the emission of a 7.68 MeV 
$\alpha$-particle 164 $\mu$s later.
The main remaining background expected are neutrons, especially 
because of the large cross section for neutron capture on \cdhdz , 
producing gammas up to 8 MeV. However, a severe neutron shield or
an isotopical derichment of \cdhdz{} reduces the background significantly.
Another background source is possible 
radio-isotopes produced by cosmic ray spallation. They were measured
for the $\gamma$-astronomy mission INTEGRAL in a proton test beam at 
CERN \cite{Por96} and only two dangerous 
isotopes are identified. Fortunately they are all rather short-living and have a low
branching ratio.\\
Last but not least, the tail of \zbb{} leaks into the peak region
of \obb, because it occurs with an orders of magnitude higher rate. The crucial
quantity to keep this under control is the energy resolution of the
detector. A very promising choice is a semiconductor like CdTe. For a  To improve
the neutrino mass further the experiment has to be upgraded to larger masses, which
according to the modular design can be performed rather easily.

\section{Acknowledgements}
We thank C.~G\"o{\ss}ling for useful discussions and
his support. We also acknowledge the
support of Th. Villett and the mechanical workshop of the University of Dortmund
during construction of the test setup.  K.~Zuber is supported by a
Heisenberg Fellowship of the Deutsche Forschungsgemeinschaft. 
Finally we want to thank eV-PRODUCTS for lending a detector for
test measurements.

\bibliographystyle{elsart-num}
\bibliography{rare}

\end{document}

%% file: kai.tex
Conservation laws play a crucial role in modern particle physics,
normally resulting from invariances under certain symmetry transformations.
However, there are symmetries in the Standard Model, where such an
underlying invariance is not known, the most popular ones are baryon
and lepton number conservation. Therefore, the investigation of lepton
number violating processes is one of the most promising ways of probing
physics beyond the Standard Model. A particular aspect of this topic
is lepton number violation in the \neu\ sector, which in the case of
massive \neus\ would allow a variety of new phenomena, for a recent 
review see \cite{zuber}.

Lepton number violation emerges immediately in case of the existence
of \majo\ masses of the neutrino, which is predicted in most
GUT-theories, by using the see-saw mechanism to explain small neutrino
masses \cite{murray,yan,rabi}.

The gold-plated channel to probe the \majo\ character
of neutrinos is the $\Delta L = 2$ process of \onbb\ (\nbb):
\begin{equation}
(Z,A) \ra (Z+2,A) + 2 e^-
\end{equation}
The measured half-life is correlated with the neutrino mass
via
\begin{equation}
\label{hwz}
T_{1/2}^{-1} = G^{0\nu} M_{0\nu}^2 \left(\frac{\langle m_{ee} \rangle}{m_e}\right)^2
\end{equation}
with $G^{0\nu}$ as the phase space and $M_{0\nu}^2$ as the nuclear
transition matrix elements.
The quantity $\langle m_{ee} \rangle$ is called the effective Majorana
mass and is given by
\begin{equation}
\langle m_{ee} \rangle = \mid \sum U_{ei}^2 m_i \eta^{CP} \mid 
\end{equation}
where $m_i$ are the mass eigenstates, $\eta^{CP} = \pm 1$ the relative
CP-phases and $U_{ei}$ the mixing matrix elements. 
The other eight possible effective Majorana masses $\langle m_{\alpha\beta} \rangle$
with $\alpha, \beta = e, \mu, \tau$ are currently not restricted at all
\cite{zub02}.

\begin{figure}[htb]
\centering
\includegraphics[width=.9\linewidth]{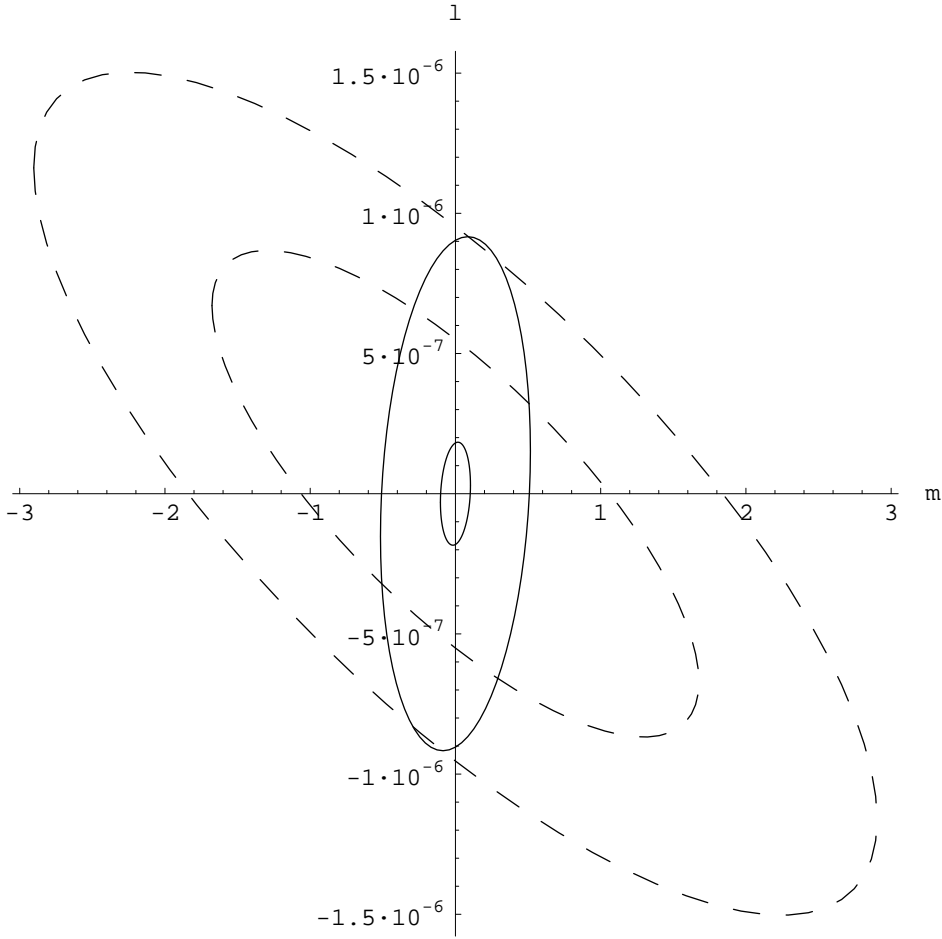}
\caption{Complementary of \obb{} and \bpbp{}-decay modes. Shown is the $\langle \lambda \rangle$ - \ema
plane (assuming $\langle \eta \rangle=0$ for simplicity). The small ellipse (solid line) corresponds
to the allowed parameter values of the recently claimed evidence for \obb{} \protect \cite{evidence} (see also
\protect \cite{critics}). Also
shown are the parameter regions (dashed lines) corresponding to a possible \bpbp-measurement with a half-life
between $1 \cdot 10^{26}$ and $3 \cdot 10^{26}$ years. 
\label{ellips}}
\end{figure}

In addition, the Standard Model process of \zbb{} can occur
\begin{equation}
(Z,A) \ra (Z+2,A) + 2 e^- + 2 \bar{\nu}_e
\end{equation}
whose detection is important to check the reliability of nuclear matrix
element calculations.
Various other mechanisms beside a light Majorana neutrino exchange have
been proposed to mediate \obb, among
them are \eg\ right-handed weak currents \cite{doi} and R-parity violating SUSY \cite{moha1,hirsch}. To
disentangle the underlying mechanism it is worthwhile to explore additional processes. 
In case of the existence of right handed weak currents the full Hamiltonian for double beta decay
can be written as
\begin{equation}
H \sim  (j_L J_L^\dag + \kappa j_L J_R^\dag +
\eta j_R J_L^\dag + \lambda j_R J_R^\dag)
\end{equation}
where $j$ denoting leptonic currents and $J$ hadronic
currents and $\kappa,\eta,\lambda \ll 1$. The subscript L denotes
left-handed weak charged currents of (V-A) structure and R corresponds
to right-handed weak charged currents of (V+A) type, which have not
been observed yet. 
In this case eq. \ref{hwz} has to be replaced by
\begin{eqnarray}
\label{eq:cmm} T_{1/2}^{-1} & = & C_{mm} (\frac{\ema}{m_e})^2 + C_{\eta \eta} \langle \eta
\rangle^2 + C_{\lambda \lambda} \langle \lambda \rangle ^2 \\
& & + C_{m\eta}(\frac{\ema}{m_e})\langle \eta \rangle + C_{m\lambda}(\frac{\ema}{m_e})\langle
\lambda\rangle + C_{\eta\lambda}\langle \eta \rangle \langle \lambda \rangle
\end{eqnarray}
where the coefficients $C$ contain the phase space factors and the matrix elements and
\begin{equation}
\langle \eta \rangle = \eta \sum_j U_{ej}V_{ej} \quad \langle \lambda \rangle = \lambda \sum_j
U_{ej}V_{ej}
\end{equation}
$V_{ej}$ correspond to the mixing matrix elements among the right-handed neutrino states.
This results in an ellipsoid in the three parameters \ema, $\langle \lambda \rangle$ and $\langle
\eta \rangle$. Decay modes which are dominantly driven
by right-handed weak currents are  transitions to excited states and \bpbp-decays.
The latter can occur in three variants: 
\begin{eqnarray}
          (Z,A) &\ra& (Z-2,A) + 2 e^+(+ 2 \nel) \quad\quad\,\, \mbox{(\bpbp{})} \label{eqbb}\\
  e_B^- + (Z,A) &\ra& (Z-2,A) + e^+ (+ 2 \nel)  \quad\quad \mbox{(\bec{})} \label{eqbe}\\
2 e_B^- + (Z,A) &\ra& (Z-2,A) (+ 2 \nel)        \quad\quad\quad\quad \mbox{(\ecec{})} \label{eqee}
\end{eqnarray}
Like in \nbb\ the processes (\ref{eqbb})-(\ref{eqee}) can occur with and without
the emission of neutrinos. The Q-value of the transition is reduced by $m_ec^2$ for any
produced positrons. Therefore, \ecec{} has the largest phase space.
Clearly neutrinoless \ecec{} in the stated form violates energy 
and momentum conservation, therefore additional particles have to be emitted. As discussed in 
\cite{doi93} the most likely mechanism is the creation of an additional electron-positron pair
via internal conversion. How both modes might disentangle the underlying parameters is 
shown in \fig~\ref{ellips} following the description of \cite{hir94}.

\begin{figure}[htb]
\centering
\includegraphics[width=.9\linewidth]{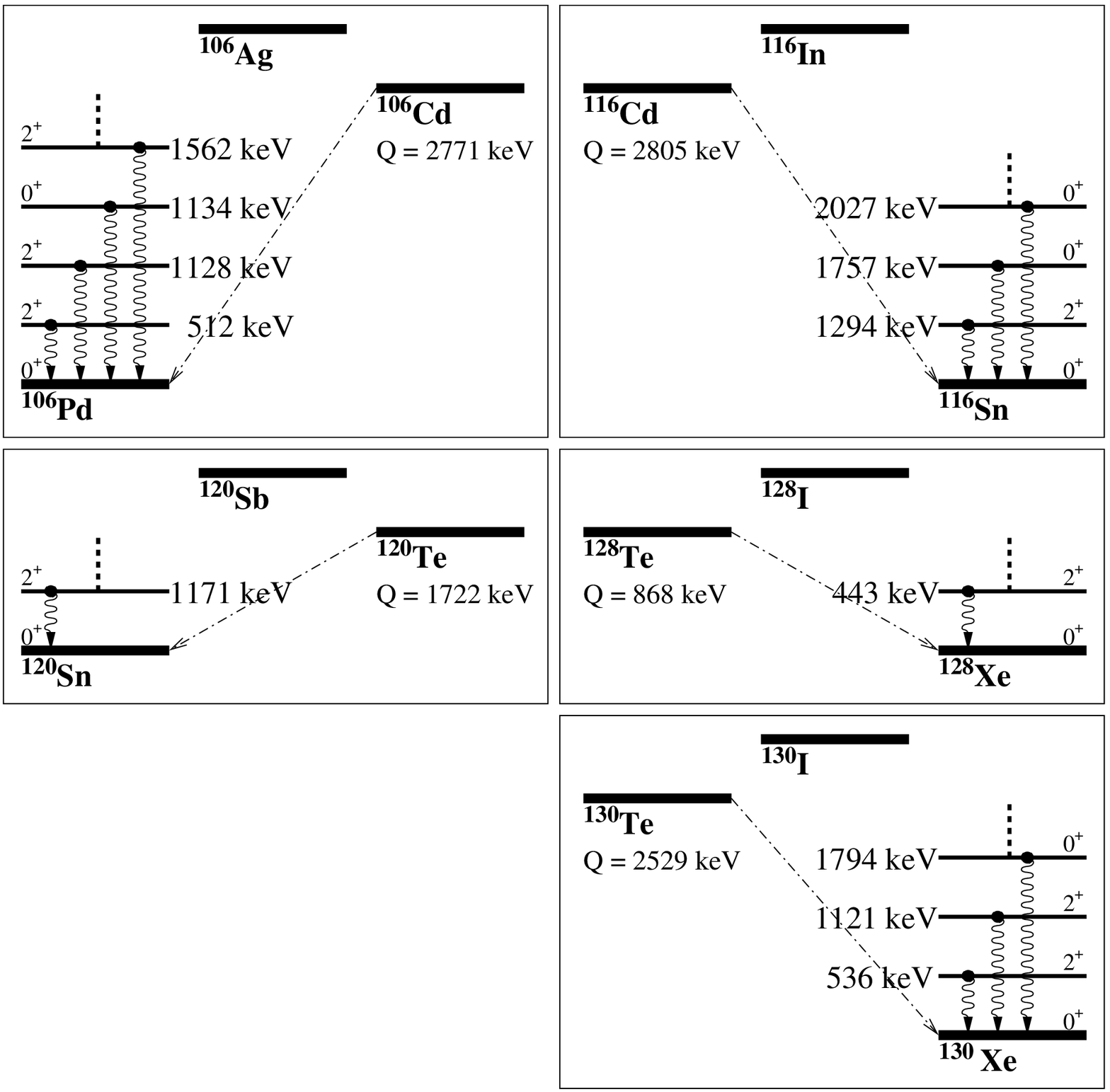}
\caption{Decay schemes of the various double beta transitions to excited
	states.  The energies of all investigated $J^\pi= 0^+$ and $2^+$
	states are given with their respective energy levels. Left:
	\bpbp-decay, right: \bmbm-decay. Top: \cdhs\ and \cd, middle:
	\tehz\ and \tehaz, bottom: \tehd.
\label{levels}}
\end{figure}
 
In this paper we explore the various decay channels of Cd,
Te and
Zn isotopes.  This is done by using CdZnTe and CdTe semiconductor
detectors as a first step towards the realisation of the COBRA-project
\cite{COBRA}. The corresponding decay schemes of all investigated isotopes
are shown in \fig~\ref{levels}, taken from \cite{toi}.
For a recent review on double beta decay see \cite{steve}
and a compilation of current results can be found in
\cite{Tretyak,Tretyak2}.

%% file: setup.tex
\section{Experimental setup}
This analysis is based on measurements performed with two detectors,
a 2.89\,g $10 \times 10 \times 5\,\mbox{mm}^3$ CdZnTe detector provided
by eV-PRODUCTS and a 5.8\,g $10 \times 10 \times 10\,\mbox{mm}^3$ CdTe
detector purchased from EURORAD.  Both are standard industrial products.
The first detector (CPG) and its
preamplifier are encapsulated within an aluminium tube with a 2\,cm
tungsten shield in between. The second detector (ER) is situated in
a copper case with the preamplifier outside of a passive shielding.

The ER detector is read out conventionally and shows asymmetric photo
peaks with a typical tail towards lower energies due to insufficient
collection of produced holes caused by trapping, whereas for the CPG
detector the readout is based on coplanar grid technology.  This
technique allows to read out the pure electron signal only resulting
in almost symmetric photo peaks \cite{cpgpaper}.

\begin{figure}[htb]
\centering
\includegraphics[width=.9\linewidth]{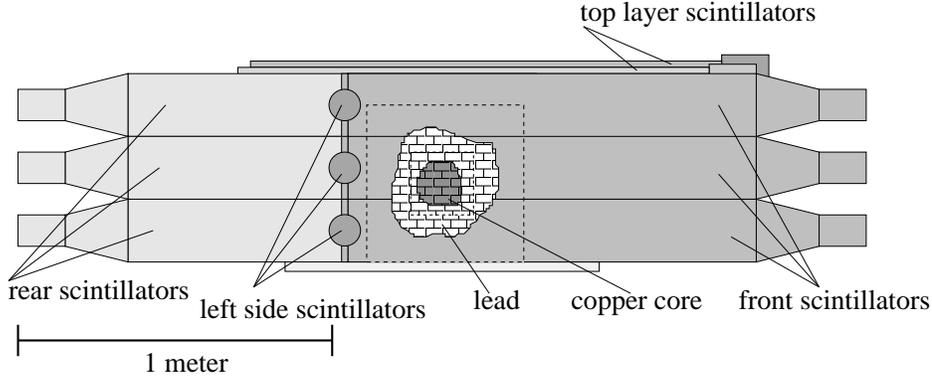}
\caption{Schematic drawing of the used setup showing the
	active and passive shielding components.
\label{setupfig}}
\end{figure}

To explore decays with extremely low expected decay rates, certain
prerequisites have to be taken to reduce the background to a minimum.
To achieve this, a special shielding (\fig~\ref{setupfig}) has been
set up to reduce background coming from cosmic ray muons and
environmental radioactivity.
This shielding consists of passive and active components.  The
passive part is a $50 \times 50 \times 50\,\mbox{cm}^3$ casing made
of copper and lead where the inner layer around the detectors consists
of electrolytic copper and the outer layer of spectroscopy lead.  The
lead bricks have been cleaned in an ultrasonic acetone bath
and their surfaces have been etched with HCl and HNO${}_3$ before
assembly. This 1.4\,t passive shielding is covered with an airtight
$60 \times 60 \times 60\,\mbox{cm}^3$ aluminium box constantly flushed
with nitrogen to protect the apparatus from $^{222}$Rn.
The detectors themselves are located in a specially manufactured
copper brick.

The active part of the shielding against cosmic rays consists of
19 plastic scintillators of $130 \times 20 \times 0.5\,\mbox{cm}^3$
mounted on the aluminium box.  The four sides are covered with one
layer of three scintillators each and the top with two layers of
four and three scintillators displaced by 50\% of their width to
minimize dead areas. All photomultiplier signals are fed into
discriminators whose OR output signal triggers a 20\,$\mu$s long
veto.  Running at a trigger rate of about 800\,Hz the resulting
dead time is 1.65\%.  The detection efficiency of cosmic ray muons
is 95\% for a single layer and about 99.7\% for the double layer.

The whole apparatus is located in a building with barite concrete
walls and ceilings resulting in a total shielding of about 5\,mwe. 

The data acquisition is done with a 13bit PC multichannel analyser
(MCA).  The pre-amplified signals of the detectors are fed into
shaping amplifiers with a shaping time of $0.5\,\mu$s.  The gain
of the amplifiers is adjusted to produce a signal of approximately
10\,V at 4\,MeV, which is the maximal allowed input voltage of the
MCA's ADC.  The lower threshold is chosen such that the dead time
of the ADC did not exceed 2\%.

\begin{figure}[htb]
\centering
\includegraphics[width=.9\linewidth]{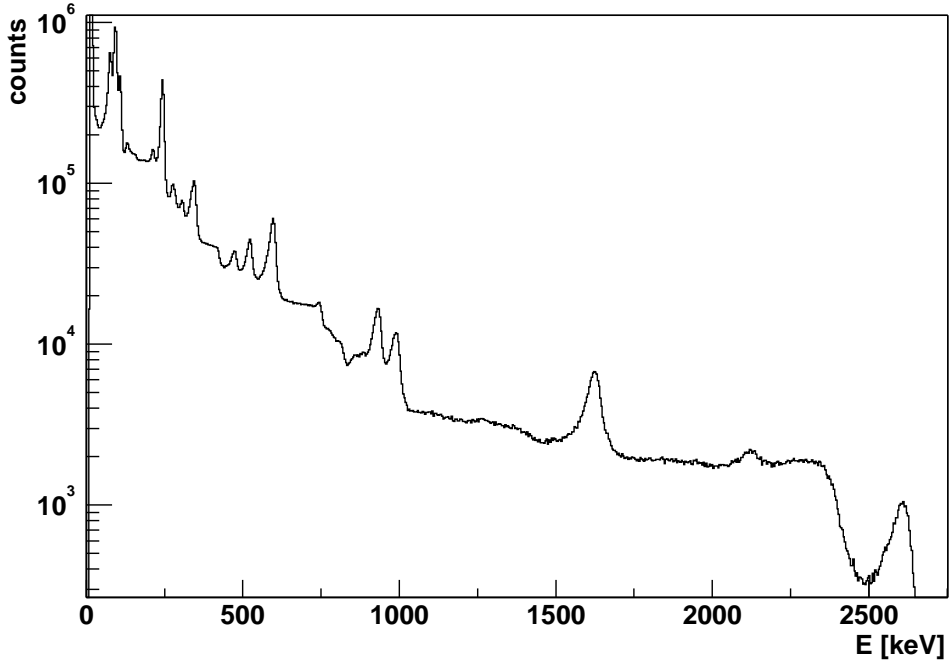}
\caption{$^{228}$Th spectrum as obtained with the CPG detector.
	The most prominent lines at 238.6\,keV, 583.19\,keV,
	911.21\,keV, 968.97\,keV, 1592\,keV and 2614\,keV are
	clearly visible. The $^{208}$Tl line at 2614\,keV serves
	as an important calibration point because of its close
	location to the \obb\ regions of the \cd\ and \tehd\ decays.
\label{thcalib}}
\end{figure}

The energy resolution of the detectors was measured with various
sources, among them $^{241}$Am, $^{57}$Co, $^{137}$Cs, $^{60}$Co
and $^{228}$Th (\fig~\ref{thcalib}) and is assumed to rise with
the square root of the energy within the region of interest.  The
resulting FWHM energy resolution for the CPG detector is
$ \sDelta E_{\mbox{\tiny CPG}} = \sqrt{E/\mbox{keV}} \times 1.03\,\mbox{keV} - 0.72\,\mbox{keV} $
and for the ER detector
$ \sDelta E_{\mbox{\tiny ER}}  = \sqrt{E/\mbox{keV}} \times 1.68\,\mbox{keV} + 3.66\,\mbox{keV}\,. $
The sources were also used for regular calibrations during data
taking to assure stable operation of the detectors.
Hence, it could be shown that the variation of the peak positions is well below 0.5\,\%.

%% file: data.tex
\section{Data Reduction}
\begin{figure}[htb]
\centering
\includegraphics[width=.9\linewidth]{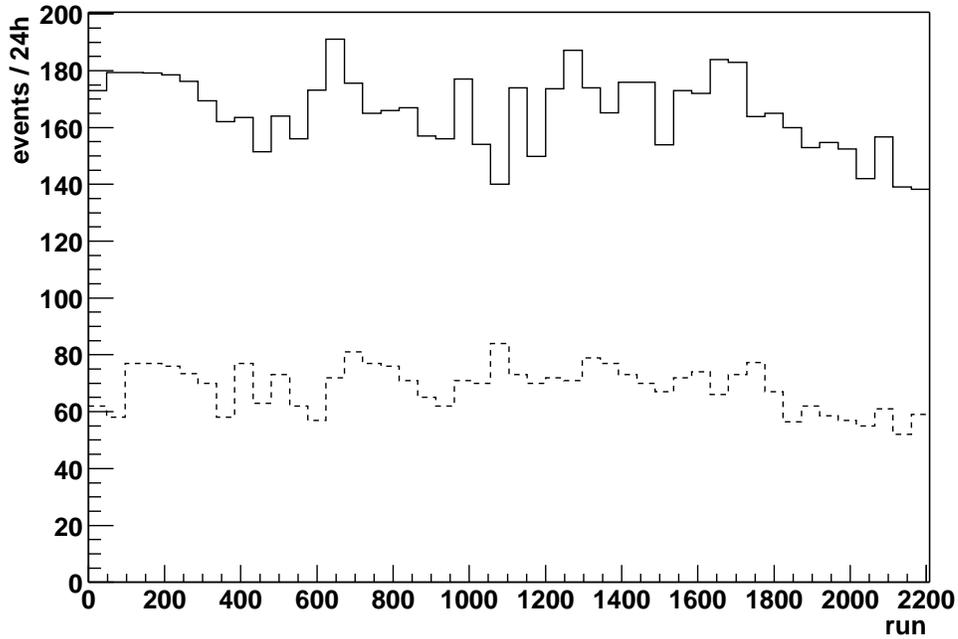}
\caption{Events per day for recorded CPG data in the intervals
	150-300\,keV (solid line), 0.5-2\,MeV (dashed line),
	confirming stable operation of the detector. The run
	period shown corresponds to about 46.5\,days.
\label{cpgtime}}
\end{figure}

\begin{figure}[htb]
\centering
\includegraphics[width=.9\linewidth]{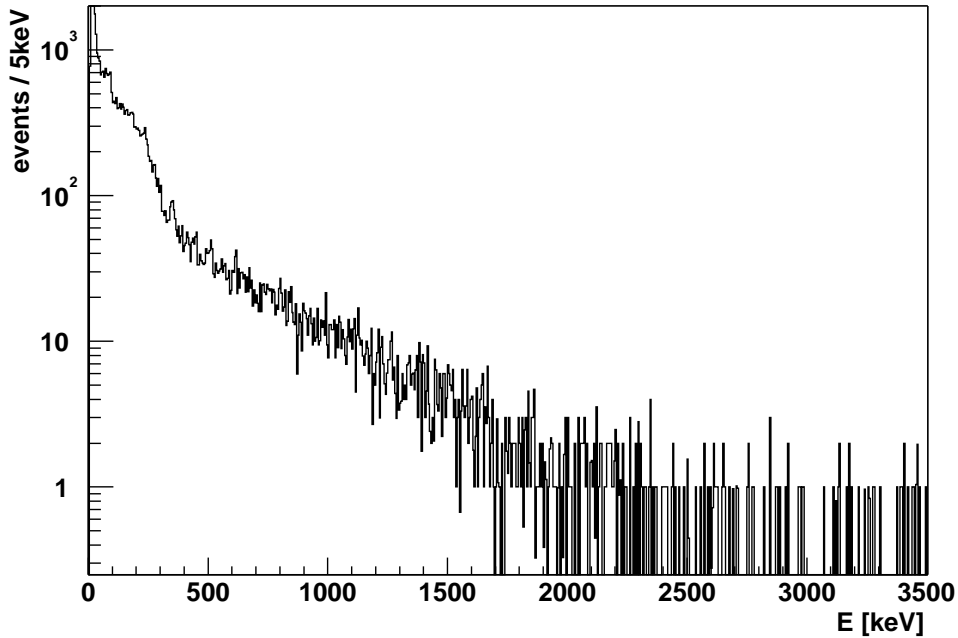}
\caption{Measured CPG spectrum up to 3.5\,MeV for 134.5\,g$\cdot$d
	in events per 5\,keV. Small peaks of the 351\,keV and 609\,keV lines
	of the $^{238}$U decay chain can be seen. A shoulder at low energies 
	due to the fourfold forbidden \bdec\ of \cdhdz\ is clearly
	visible.}
\label{spectrumcpg}
\end{figure}

During the data taking phase runs of 30 minutes life-time have been
recorded.  The distribution of obtained single runs follows nicely
a Poisson distribution, however, there are a few bursts, probably
due to electric disturbances.
Such runs were discarded, in case of the CPG detector there were
two such burst runs.  The remaining runs show a stable event rate
for the intervals 150\,keV\,-\,300\,keV and 0.5\,MeV\,-\,2\,MeV
(\fig~\ref{cpgtime}) leading to a total measuring time of 1117
hours.  The rebinned final spectrum used for the analysis is shown
in \fig~\ref{spectrumcpg}.


The spectrum shows several features of background sources.  The most
prominent signature is a bump below 300\,keV which can be attributed
to the fourfold forbidden non-unique beta spectrum of \cdhdz.
Also visible are small peaks of the ${}^{238}$U (351\,keV and 609\,keV) and
${}^{232}$Th (238.6\,keV and 338\,keV) decay chains.

For the ER detector also runs of 30 minutes length have been recorded.
However, in these runs some systematic irregularities are apparent in the
low energy range.
Therefore, below 1\,MeV a variation in the average rates could 
be observed, making the data unreliable in that region.
For the spectrum above
1\,MeV this effect does not show up, so that almost the full data set
can be used, rejecting only burst runs with exceptional high count rates
as for the CPG detector.
The usable data set corresponds to 397.5\,g$\cdot$d.

\begin{figure}[htb]
\centering
\includegraphics[width=.9\linewidth]{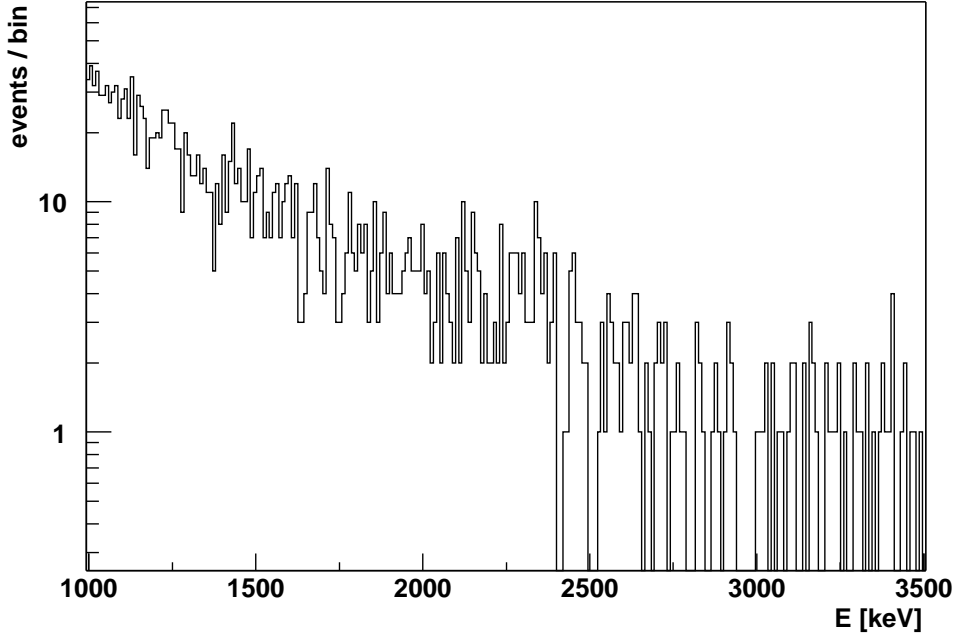}
\caption{Measured ER spectrum above 1\,MeV in events per 9.45\,keV.
	This spectrum corresponds to 397.5\,g$\cdot$d of data taking. 
\label{4MeVspectrume2}}
\end{figure}

The spectrum recorded with the ER detector as shown in
\fig~\ref{4MeVspectrume2} does not reveal clear background lines
mainly due to its worse energy resolution.  Nevertheless, some
${}^{232}$Th may be present implied by a slight enhancement of
events around 2614\,keV.

%% file: analysis.tex
\section{Data analysis}
In case the half-life $T_{1/2}$ under investigation exceeds by far
the measuring time ($t\ll T_{1/2})$ the radioactive decay law 
including the efficiency $\epsilon$ to detect the event is given by
\begin{equation}
T_{1/2} = \frac{N_0 \cdot \ln\!2 \cdot t \cdot \epsilon}{n}
\end{equation}
where $N_0$ is the number of atoms of a particular isotope in
the source, $t$ the time of measurement and $n$ the number of
observed events, or for limits the number of excluded
events.

\begin{figure}[htb]
\centering
\includegraphics[width=.9\linewidth]{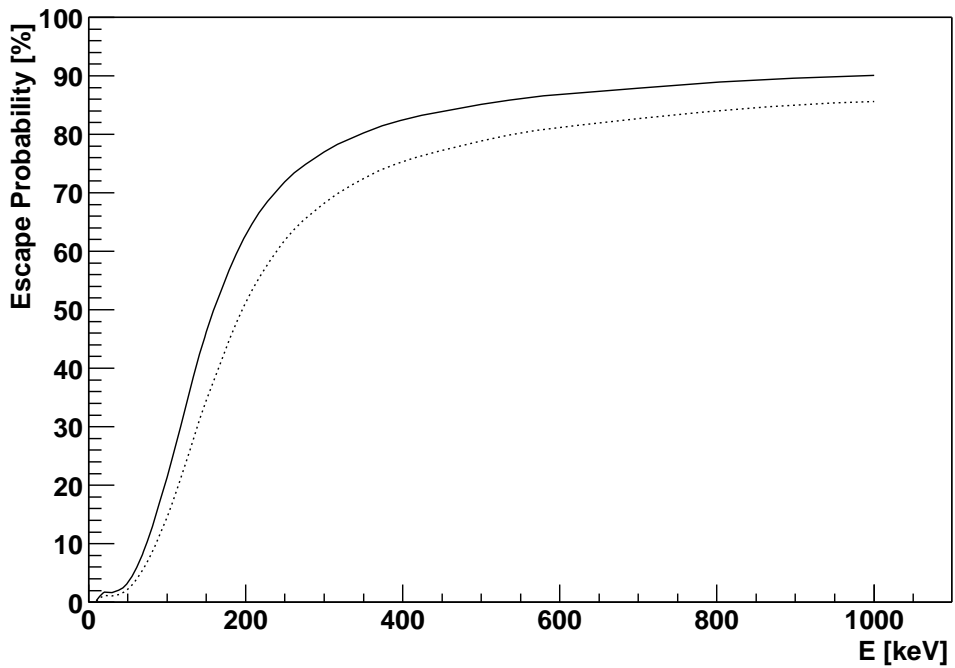}
\caption{Probability of a gamma to escape from a detector without energy
	deposition as a function of its energy as calculated with GEANT4.
	The solid line represents the CPG, the dotted line the ER detector.
\label{escapefig}}
\end{figure}

The efficiencies for both detectors have been calculated with the Monte Carlo
package GEANT4 \cite{GEANT} and cross checked with the MCNP program package \cite{MCNP}.
Both have been modified to generate double beta events via a neutrino
mass term where the
energy and angular distributions of the two electrons are simulated
according to formulae given in \cite{Tretyak}.  All events are started homogenously and
isotropically in the whole detector.  The simulations calculate the
energy deposition of $e^-$, $e^+$ and $\gamma$ from an event 
within the detector including Compton backscattering from the shielding.
Both programs produce consistent efficiencies.
As an example, the efficiencies for full absorption (resp.~no energy
deposition) of 511\,keV annihilation gammas is 4.5\%\,(85.0\%) and
7.6\%\,(79.2\%) for the CPG and ER detectors.  The
efficiencies for no energy deposition of photons less than 1\,MeV
are shown in \fig~\ref{escapefig}.

The natural abundances, decay modes and \qvals\ of all isotopes under study are
given in \tab~\ref{naturalisotopes}.  The composition of the CPG
detector is given by $(\mbox{Cd}_{0.9}\,\mbox{Zn}_{0.1}\mbox{Te})$
with a density of 5.78\,$\mbox{g}/\mbox{cm}^3$.  The density of
the ER CdTe detector is 5.85\,$\mbox{g}/\mbox{cm}^3$.

\begin{table}
\centering
\begin{tabular}{rcrr}
Isotope & decay mode & \qval\ $[$\keV$]$ & nat. abund. $[\%]$ \\
\hline
  ${}^{64}$Zn & \bec, \ecec	& 1096 & 48.6 \\
  ${}^{70}$Zn & \bmbm		& 1001 &  0.6 \\
 ${}^{106}$Cd & \bpbp, \bec, \ecec & 2771 &  1.25 \\
 ${}^{108}$Cd & \ecec		&  231 &  0.89 \\
 ${}^{114}$Cd & \bmbm		&  534 & 28.72 \\
 ${}^{116}$Cd & \bmbm		& 2805 &  7.47 \\
 ${}^{120}$Te & \bec, \ecec	& 1722 &  0.096 \\
 ${}^{128}$Te & \bmbm		&  868 & 31.69 \\
 ${}^{130}$Te & \bmbm		& 2529 & 33.80 \\
\end{tabular}
\caption{Decay modes, \qvals\ and natural abundances of isotopes under study.
\label{naturalisotopes}}
\end{table}

In this analysis the lower limits of half-lives are obtained via
the following $\chi^2$-method:  The background around the position of
the peak under investigation is fitted with an exponential function.
A gaussian peak is then grown on the resulting function at the
energy of the searched peak and with a width as described by the
corresponding energy resolution of the detector.  The $\chi^2(n)$
of this model function is calculated as a function of the number $n$
of events in this peak, with which the null-hypothesis is checked.
The obtained $\chi^2$ can then be transformed into a function of
the inverse half-life.  To produce a joint limit for many spectra
(in this case two) the $\chi^2((T_{1/2})^{-1})$ of the single detectors
are
added.  The 90\%\,CL lower limit can then be extracted by taking
the half-life where $\chi^2_{sum}((T_{1/2})^{-1})$ is increased
by 2.71 with respect to its minimum.

%% file: modes.tex
\section{Results}
A variety of signatures of the different double beta decay modes
are investigated.  In all cases only neutrinoless modes are
considered, except for double electron capture (\ecec) where
the neutrino accompanied mode is used.

The following limits are calculated according to the above 
$\chi^2$-method from 1117\,h measurement with a 2.89\,g CdZnTe detector
(134.5\,g$\cdot$\,d) and additionally from 1645h with a 5.8\,g CdTe
detector (397.5\,g$\cdot$d) for energies above 1\,MeV.

\subsection{\bmbm--transitions}
\begin{figure}[htb]
\centering
\includegraphics[width=.9\linewidth]{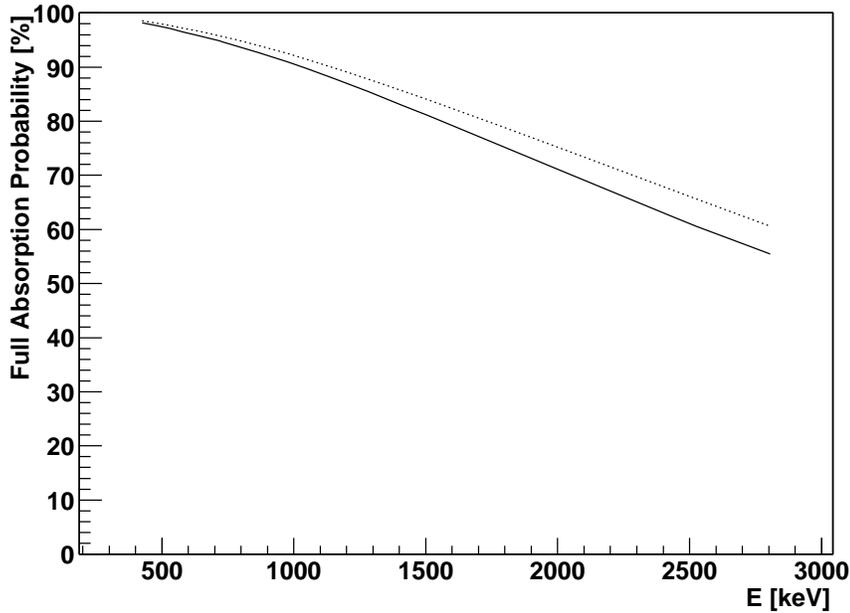}
\caption{Probability to detect the full energy of both electrons of
	a \nbb-decay as calculated with GEANT4.  The solid line
	represents the CPG, the dotted line the ER detector.
\label{doubleefffig}}
\end{figure}

The signal of the \nbb-decay to the ground state of the daughter
nucleus is a peak at the \qval\ of the involved nuclear transition, which
is given in \tab~\ref{naturalisotopes}.
Due to the fact that source and detector are identical the
detection efficiency is high for this decay mode as can be seen
from \fig~\ref{doubleefffig}.  Typical peak efficiencies assuming the
neutrino mass mechanism are
55.5\%\,(60.6\%) for \cd\ ($Q=2.805$\,MeV), increasing to
92.7\%\,(94.0\%) for \tehaz\ ($Q=868$\,keV) for the CPG (ER)
detector. In case of right handed weak currents the efficency of the
CPG detector is slightly less, about 53\%.

The isotope with the highest \qval\ is \cd.  The background
in the \cd, where the exponential function is basically flat,
can be estimated by averaging the spectrum from 2.7 to 3.4\,MeV
while excluding the peak region at 2.8\,MeV to be
$3.3\times{10}^{-4}\,[\mbox{keV$\cdot$g$\cdot$d}]^{-1}$.

\begin{figure}[htb]
\centering
\includegraphics[width=.9\linewidth]{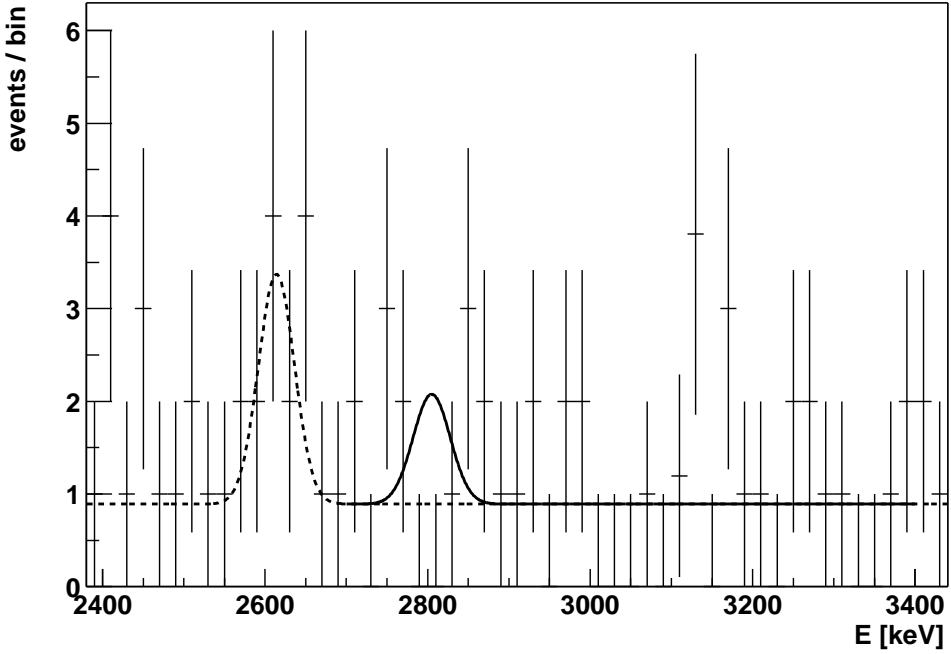}
\caption{Fit of constant background plus 2614\,keV ${}^{208}$Tl line
	(dashed line) with $\chi^2/\mbox{ndf} = 18.3/30$ and exclusion
	plot for 2805\,keV \cd\ \nbb\ (solid line)
	using the rebinned CPG spectrum.
\label{cpg_cd}}
\end{figure}

\begin{figure}[htb]
\centering
\includegraphics[width=.9\linewidth]{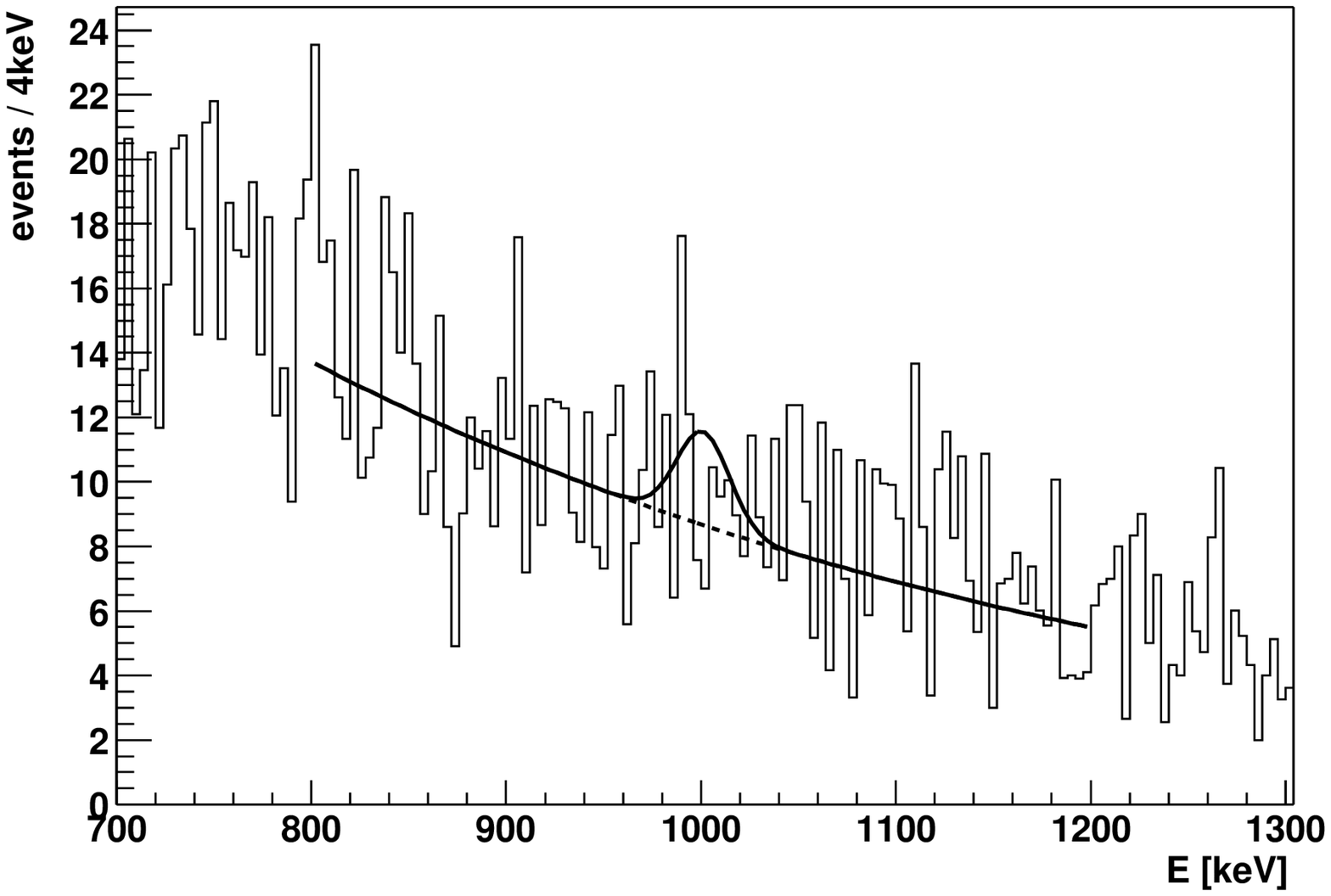}
\caption{Fit of exponential background (dashed line) from 800-1200\,keV
	with $\chi^2/\mbox{ndf} = 213.8/99$ and exclusion plot for
	1001\,keV ${}^{70}$Zn \nbb\ (solid line) using the
	rebinned CPG spectrum.  Using different fit intervals
	(750-1250, 800-1200 and 850-1250\,keV), the
	number of excluded events is stable within 4\,\%.
\label{cpg_zn}}
\end{figure}

\Fig~\ref{cpg_cd} shows the region of interest for the CPG spectrum
only.  The dashed line represents the flat background plus a fitted
$^{208}$Tl line at 2614\,keV of $6.4\pm4.5$ events.  This background model
yields a $\chi^2/\mbox{ndf} = 18.3/30$.  The 90\%\,CL excluded number of
events for \cd\ \nbb-decay ($Q=2805$\,keV, 3.4
events) as obtained by the $\chi^2$-method is superimposed.
In a similar fashion all other transitions are evaluated.
Evidently, for \zns\ (\fig~\ref{cpg_zn}) only the CdZnTe detector
(CPG) could be used.  No peak is observed at any transition energy.

The search for transitions to excited states is also looking for
a peak, however, at $Q - E_\gamma$, requiring that the energetic
gammas of the de-excitation must leave the detector without any
energy deposition through Compton scattering, photo effect or
pair-production.  De-excitation photons in the range from 400\,keV
up to 2.3\,MeV are considered and are given in \tabs~\ref{limitsminus}
and \ref{limitsplus}.

A typical escape efficiency for the ER detector is 86.1\% for 
$E_\gamma=1294$\,keV (see \fig~\ref{escapefig}) coming from the
\cd\ decay into the first excited state of ${}^{116}$Sn\,($2^+$).
At all the relevant positions searches for peaks analogous to the
ground state transitions were performed.  A compilation of all
obtained results is shown in \tab~\ref{limitsminus}.

\input{results.tex}

The achieved half-life limits are of the order of $0.7\times {10}^{18}$
to $3.3\times {10}^{19}\,yrs$ for various cadmium and tellurium
\nbb-modes.  Using only a rather small amount of CdTe and CdZnTe
it cannot be expected to probe the current limits for \cd\ and
\tehd\ \cite{Tretyak2}.  Nevertheless, for \zns\ an improved limit
of $T_{1/2} > 1.3 \cdot 10^{16}\,yrs$ (90\%\,CL) is obtained.

\subsection{\bpbp--transitions}
The \bpbp-modes are more difficult to detect and have hence
lower half-life limits.  The energy available for the decay
decreases by $2 m_e c^2 = 1022$\,keV per $\beta^+$ compared with
the electron capture (\EC) modes.
The energy distribution of the positrons in a \npp-decay 
is the same as for the \nbb-mode and the energy
deposited by the two positrons is $E=Q-4m_e c^2$.
In the \npe\ mode the total energy is carried away by the single
positron only.  Additionally, the energy of the electron capture
process is added, again leading to a line at $E=Q-2m_e c^2+E_K$, because
elctron capture is dominantly from the K-shell.
2$\nu$\ecec\ is only detectable via ``soft'' x-ray cascades or Auger
electrons with a total energy of about two times the binding energy
of the K-shell, assuming that K-capture is the major contribution to
\EC. Both possibilities can be detected equally well, because all
the energy will be deposited in the CdTe detector.  The \ecec\ detection
requires
a rather low energy threshold. Only the CPG detector was used for that,
running with a threshold of about 20\,keV.

In previous searches mostly 511\,keV annihilation gammas from
the positrons were detected by external detectors. In our case also the
positron energy
can be measured to discriminate different isotopes.
\ecec\ can be investigated as well.

Like in the \bmbm-case, also here searches for excited
state transitions can be performed.  For the \npp\ and \npe\ decays
to excited states a line is searched at $E=Q-4m_e c^2-E_\gamma$ and
$E=Q-2m_e c^2+E_K-E_\gamma$ respectively.  

However, to distinguish \ecec\ excited modes from the decay
to the ground state, also the gamma has to be detected.
Unfortunately, the efficiency to fully detect these gammas is
quite poor (1.0\% for $E_\gamma = 1171$\,keV with the CPG
detector, see \fig~\ref{escapefig}), resulting in even lower
half-lives.  Therefore, for
an upper limit it is more efficient to consider the more likely
case that the de-excitation gamma escapes undetected (90.8\% for
$E_\gamma = 1171$\,keV with the CPG detector) and apply this
efficiency to the limit for the decay to the ground state.

\begin{figure}[htb]
\centering
\includegraphics[width=.9\linewidth]{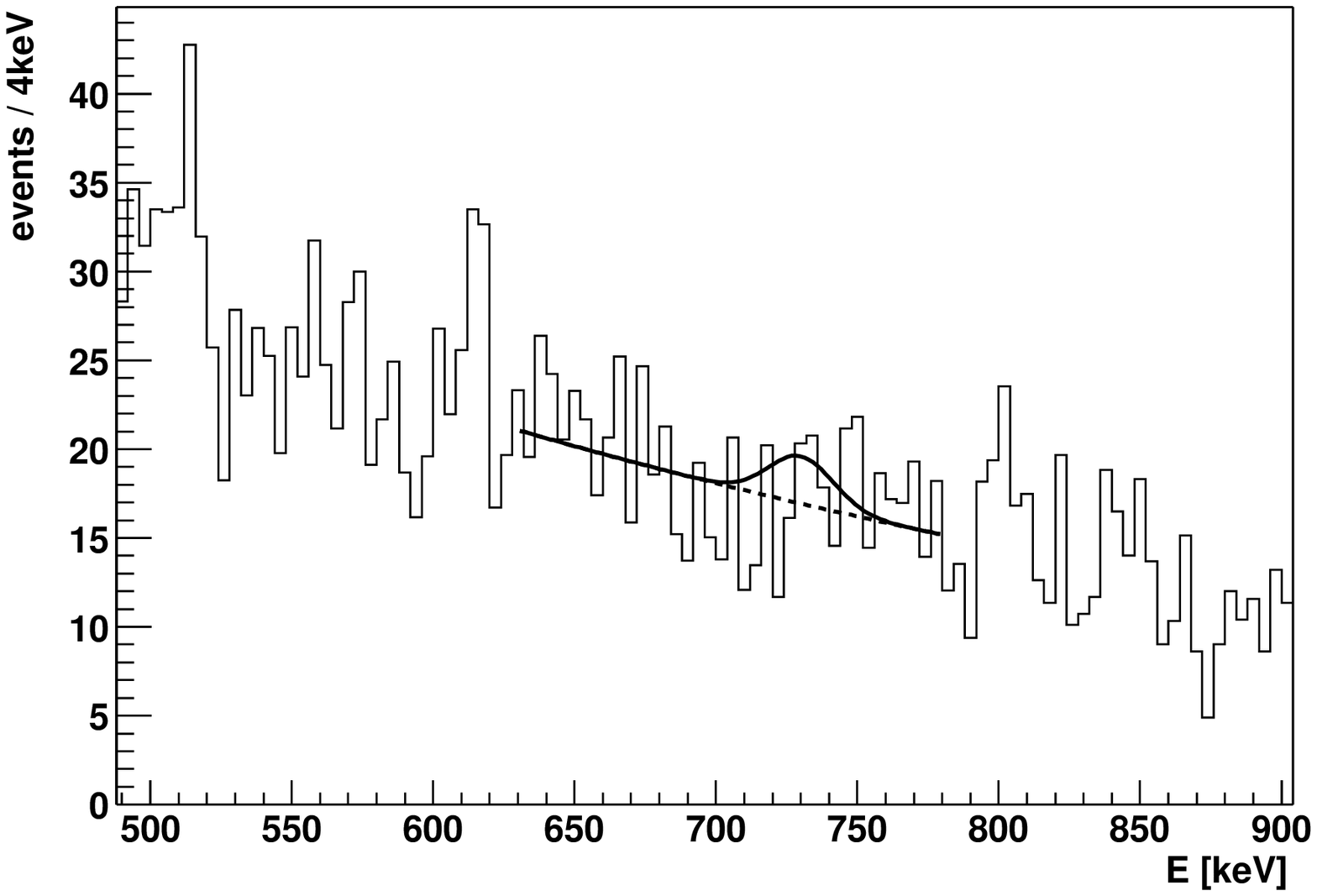}
\caption{Fit of exponential background (dashed line) from 630-780\,keV with
	$\chi^2/\mbox{ndf} = 24.6/36$ and exclusion plot for 730\,keV
	${}^{120}$Te \npe\ (solid line) using the rebinned CPG spectrum.
\label{cpg_te}}
\end{figure}

\begin{figure}[htb]
\centering
\includegraphics[width=.9\linewidth]{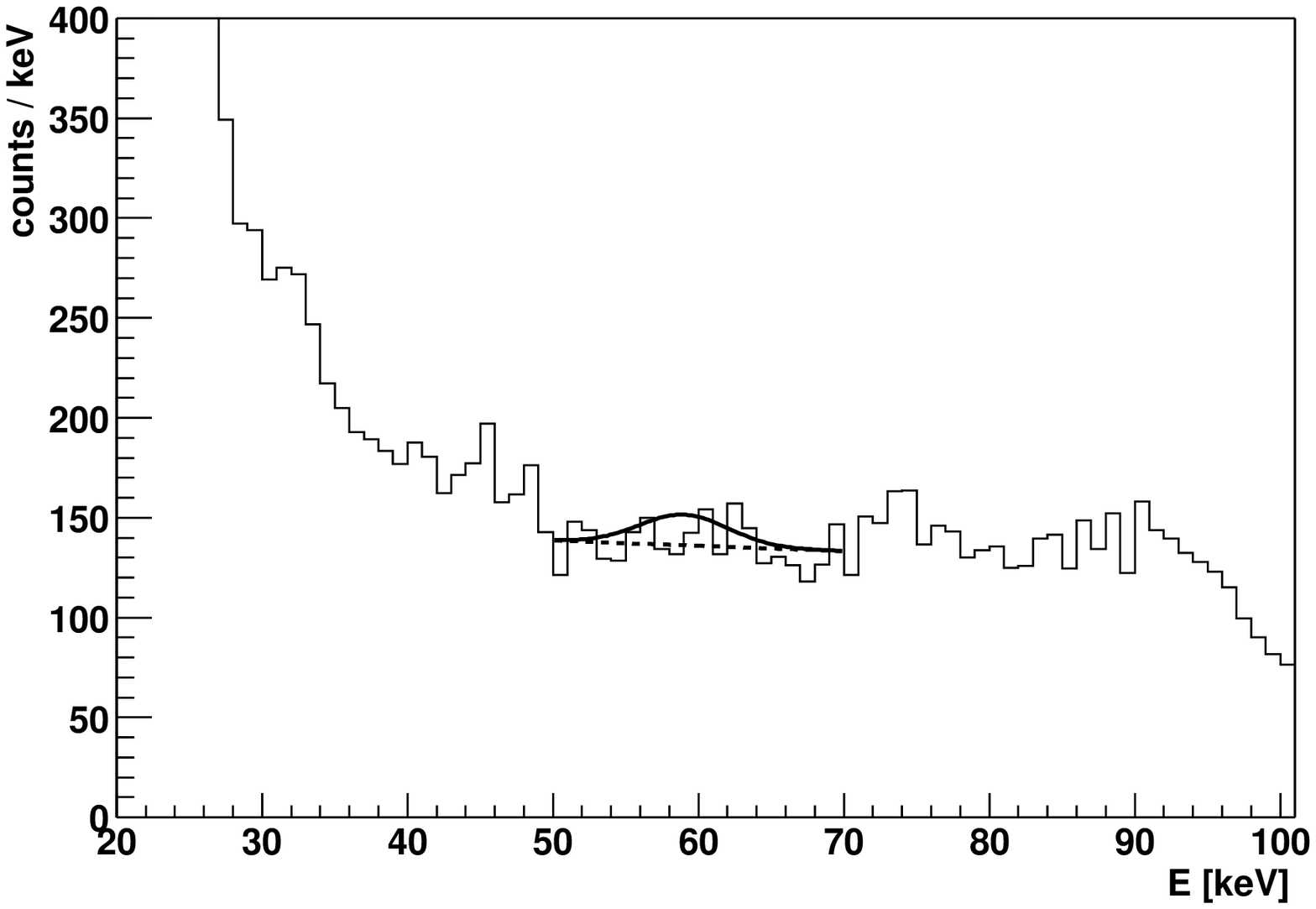}
\caption{Fit of exponential background (dashed line) from 50-70\,keV with
	$\chi^2/\mbox{ndf} = 17.1/18$ and exclusion plot for a line 
        at 59\,keV
	from ${}^{120}$Te \tee\ (solid line) using the rebinned
CPG spectrum. The peaks of the \tee\ transitions of \znvs\ and
	 \cdhs/\cdha\ lie at 17\,keV and 49\,keV respectively.
\label{cpg_teee}}
\end{figure}

\Fig~\ref{cpg_te} depicts the exclusion plot for the \npe\ decay of
\tehz.  Having a \qval\ of 1722\,keV and a K-shell binding energy of
about 30\,keV, a peak is searched at $E=Q-2m_e c^2+E_K=730$\,keV.  The
dashed line shows the fitted exponential background with
$\chi^2/\mbox{ndf} = 79.0/87$.  The solid line shows the 90\%\,CL
excluded gaussian distribution.  The same plot is shown in
\fig~\ref{cpg_teee} for the \tee-process where the signal is expected at
$2\times30$\,keV$ = $60\,keV. The corresponding energies for \znvs\ and
\cdhs/\cdha\ are 17\,keV and 49\,keV respectively.

A compilation of all obtained limits is shown in
\tab~\ref{limitsplus}.  For the various \bpbp-modes the limits are
a few orders of magnitude lower than for \bmbm-decay,
from $8.5\times {10}^{15}$ to $1.0\times {10}^{18}\,yrs$.
Some decays are investigated for the first time.
Especially for various decay modes of $^{120}$Te and
\tee\ ground state transitions of \cdhs , \cdha\ and \znvs\ new
limits could be set.

\input{results_pp.tex}

%% file: results.tex
\begin{table}
\centering
\begin{tabular}{llll}
Isotope & Level $[$\keV$]$ & $T_{1/2}$ [$yrs$] & existing limit on $T_{1/2}$ [$yrs$] \\
\hline
${}^{116}$Cd & $(0^+)$\,g.s. & $8.0\Times{10}^{18}$ & $7.0\Times {10}^{22}\, $(90\%\,CL)\cite{Dan00} \\
	     & $(2^+_1)$\,1294 & $1.6\Times{10}^{18}$ & $1.3\Times {10}^{22}\, $(90\%\,CL)\cite{Dan00} \\
	     & $(0^+_1)$\,1757 & $2.7\Times{10}^{18}$ & $7.0\Times {10}^{21}\, $(90\%\,CL)\cite{Dan00} \\
	     & $(0^+_2)$\,2027 & $7.0\Times{10}^{17}$ & $2.1\Times {10}^{21}\, $(90\%\,CL)\cite{piepke94} \\
	     & $(2^+_2)$\,2112 & $1.6\Times{10}^{18}$ & $1.7\Times {10}^{20}\, $(68\%\,CL)\cite{Bar90} \\
	     & $(2^+_3)$\,2225 & $7.1\Times{10}^{17}$ & $1.0\Times {10}^{20}\, $(68\%\,CL)\cite{Bar90} \\
\hline
${}^{114}$Cd & $(0^+)$\,g.s. & $6.4 \Times {10}^{18}$ & $2.0\Times {10}^{20}\, $(90\%\,CL)\cite{georgadze95} \\
\hline
${}^{130}$Te & $(0^+)$\,g.s. & $3.3 \Times {10}^{19}$ & $2.1\Times {10}^{23}\, $(90\%\,CL)\cite{Cre02} \\
	     & $(2^+_1)$\,536  & $1.8 \Times {10}^{19}$ & $9.7\Times {10}^{22}\, $(90\%\,CL)\cite{Ale00} \\
	     & $(2^+_2)$\,1121 & $1.4 \Times {10}^{19}$ & $2.7\Times {10}^{21}\, $(90\%\,CL)\cite{Bar01} (0+2$\nu$) \\
	     & $(0^+_1)$\,1794 & $3.1 \Times {10}^{18}$ & $2.3\Times {10}^{21}\, $(90\%\,CL)\cite{Bar01} (0+2$\nu$) \\
\hline
${}^{128}$Te & $(0^+)$\,g.s. & $8.8 \Times {10}^{18}$ & $8.6\Times {10}^{22}\, $(90\%\,CL)\cite{Ale00}\\
	     & $(2^+_1)$\,443  & $1.3 \Times {10}^{18}$ & $4.7\Times {10}^{21}\, $(68\%\,CL)\cite{bellotti87} \\
\hline
${}^{70}$Zn  & $(0^+)$\,g.s. & $1.3 \Times {10}^{16}$ & $4.8\Times {10}^{14}$ \hspace{14.0mm} \cite{Fre52} \\
\end{tabular}
\caption{Obtained lower limits (90\%\,CL) on half-lives for
         \nbb-decay modes to ground state (g.s) and excited states
	 compared with existing limits. The limit for ${}^{70}$Zn could be
         improved by more than an order of magnitude.
\label{limitsminus}}
\end{table}

%% file: results_pp.tex
\begin{table}
\centering
\begin{tabular}{lllll}
Isotope & Level $[$\keV$]$ & Mode & $T_{1/2}$\,[$yrs$] & existing limit on $T_{1/2}$\,[$yrs$] \\
\hline
${}^{120}$Te	& $(0^+)$ g.s.& \npe & $2.2\Times {10}^{16}$ & $4.2\Times {10}^{12} $\cite{Fre52} \\
		& $(0^+)$ g.s.& \tee & $9.4\Times {10}^{15}$ & --- \\
		& $(2^+_1)$ 1171& \tee & $8.4\Times {10}^{15}$ & --- \\
\hline                                  
${}^{106,108}$Cd & $(0^+)$ g.s.& \tee& $1.0\Times {10}^{18}$ & $1.5\Times {10}^{17} $(68\%\,CL)\cite{Nor84} (g.s.+512) \\
${}^{106}$Cd	& $(2^+_1)$ 512 & \tee & $8.3\Times {10}^{17}$ & $3.5\Times {10}^{18} $(90\%\,CL)\cite{Bar96c} (0+2$\nu$) \\
		& $(0^+)$ g.s.& \npp & $1.5\Times {10}^{17}$ & $2.2\Times {10}^{19} $(90\%\,CL)\cite{zpa355_433} \\
		& $(2^+_1)$ 512 & \npp & $7.4\Times {10}^{16}$ & $1.6\Times {10}^{20} $(90\%\,CL)\cite{Bel99b} (0+2$\nu$) \\
		& $(0^+)$ g.s.& \npe & $3.8\Times {10}^{17}$ & $3.7\Times {10}^{20} $(90\%\,CL)\cite{Bel99b} \\
		& $(2^+_1)$ 512 & \npe & $2.2\Times {10}^{17}$ & $2.6\Times {10}^{20} $(90\%\,CL)\cite{Bel99b} (0+2$\nu$)\\
		& $(2^+_2)$ 1128  &                            & \\
		& +$(0^+_1)$ 1134 & \npe & $7.5\Times {10}^{16}$ & --- \\
\hline
${}^{64}$Zn	& $(0^+)$ g.s.& \tee & $6.0\Times {10}^{16}$ & $8.0\Times {10}^{15} \hspace{15.9mm} $\cite{Ber53} \\
		& $(0^+)$ g.s.& \npe & $2.8\Times {10}^{16}$ & $2.3\Times {10}^{18} $(68\%\,CL)\cite{prc31_1937} \\
\end{tabular}
\caption{Lower limits (90\%\,CL) for \npp, \npe\ and \tee-decay to
	ground state (g.s) and excited states obtained from the COBRA test
	measurements compared with existing limits.
	For the transitions of \cdhs\ to 1128\,keV $(2^+_2)$ and
	1134\,keV $(0^+_1)$ only a joint limit can be given because
	of the finite energy resolution.
\label{limitsplus}}
\end{table}